# Turning Bitcoins into the Best-coins

Siddharth Rao
(Under the guidance of Dr. Vitaly Skachek)
Department of Mathematics and Informatics
University of Tartu, Estonia
<u>siddharth.rao@aalto.fi</u>

**Abstract:**

In this paper we discuss Bitcoin, the leader among the existing cryptocurrencies, to analyse its trends, success factors, current challenges and probable solutions to make it even better. In the introduction section, we discuss the history and working mechanism of Bitcoin. In the background section, we develop the ideas that evolved in the process of making a stable cryptocurrency. We also analyze the survey matrices of the present day cryptocurrencies. This survey clearly shows that Bitcoin is the clear winner among its kind. Section 3 is about the success factors of Bitcoin and the proceeding sections are a discussion about current challenges which pose as hurdles in making Bitcoin a better currency in the digital world. We finally discuss the balance between anonymity and reduced trust in the cryptocurrency world, before concluding the survey.

## 1. Introduction

### 1.1 History of Bitcoin

Bitcoin is an emerging digital - decentralized currency system which was first found in the first self published paper [1] in October 2008 by a person with a pseudonym "Satashi Nakamoto". It was launched on 11$^{th}$ of January,2009 as open-source project on Sourceforge and announced on January 3rd 2009, on the Cryptography mailing list. Since the creation of block chaining, Bitcoin gained considerable number of users and attention from the media. Many big companies like Virgin Gigantic (Spaceflight company), WordPress, PirateBay, PayPal and Ebay accept Bitcoins besides regular currency payments [2]. At the moment there are more than 150 such companies which are accepting this digital cryptocurrency as a payment for their services [3]. Most of the exchanges, among the existing 75+ digital currency exchanges, around the globe [4], offers exchange currencies service with major currencies (such as EUR, USD, GBP, JPY, SEK, AUD, etc.). At the time of writing this report (30th April, 2014), there are around 12 Million Bitcoins in circulation with the market capitalization being approximately 5.6 Billion USD, where each Bitcoin is worth $442.9 USD.

The topic of concern in the research field about Bitcoin-like currencies is that the transactions, in the Bitcoin system, are shared by the whole network without encryption. Due to its distribution feature, history of all transactions is public and available to all nodes in the peer-to-peer network which leads to privacy leakage. Bitcoin has experienced its epic rise and fall even after being the clear winner among all the existing digital currencies. In spite of three decades of research on digital currency, Bitcoin – not being a near to perfect system, has gained its publicity, in being the market leader. In this paper we analyze the factors for Bitcoin long lasting success, what has affected the Bitcoin market (factors causing fluctuations in Bitcoin rate), existing challenges to make Bitcoin, better and probable solutions to these challenges.

### 1.2 Working Mechanism of Bitcoin

The key advantage of Bitcoin system, compared to the digital currency methodologies developed before Bitcoin, is that it allows transaction between any two parties without the intervention of an authority like central bank. An open/public ledger of transactions is maintained in the peer-to-peer network of Bitcoin users. The transactions such as the number of Bitcoins received and sent by a particular user (Bitcoin address) are maintained in that ledger. The cryptographic keys which are used to sign the transactions are the basis of these Bitcoin addresses.

*Miners* in the Bitcoin system are the entities that solve the computational proof-of-work cryptographic problems and hence create the blocks which form the Bitcoin transactions. Miners are people using Bitcoin mining software or dedicated special hardware to solve the cryptographic problems using brute-force methods by exploiting the software (CPU cycles or GPU time) or hardware resources. Every such new block created by the miner is verified by the one who creates it. Each of those blocks contains a reference to previous blocks thus they form a block chain. By extending the block chain, the miner attests that he has accepted all previous blocks in thechain. Miners also get rewarded in terms of Bitcoins for solving the proof-of-work cryptographic problems.

The *blocks* in the Bitcoin system contain the list of all executed transactions and *block chains* acts as an official record removing the need for a centralized trusted authority. Every transaction in the Bitcoin system becomes effective once they have been recorded by the block and approved by the Bitcoin network. Each block comprises of three components namely: (1) Hash Value of the Previous block (2) Recent Transactions (3) A random number (Nonce) to produce the next block. The hash value of previous block helps to link the current block with the previous ones and the random nonce is used for proof-of-work system. Blocks linked in this manner in a chronological order form the block chain of the system.

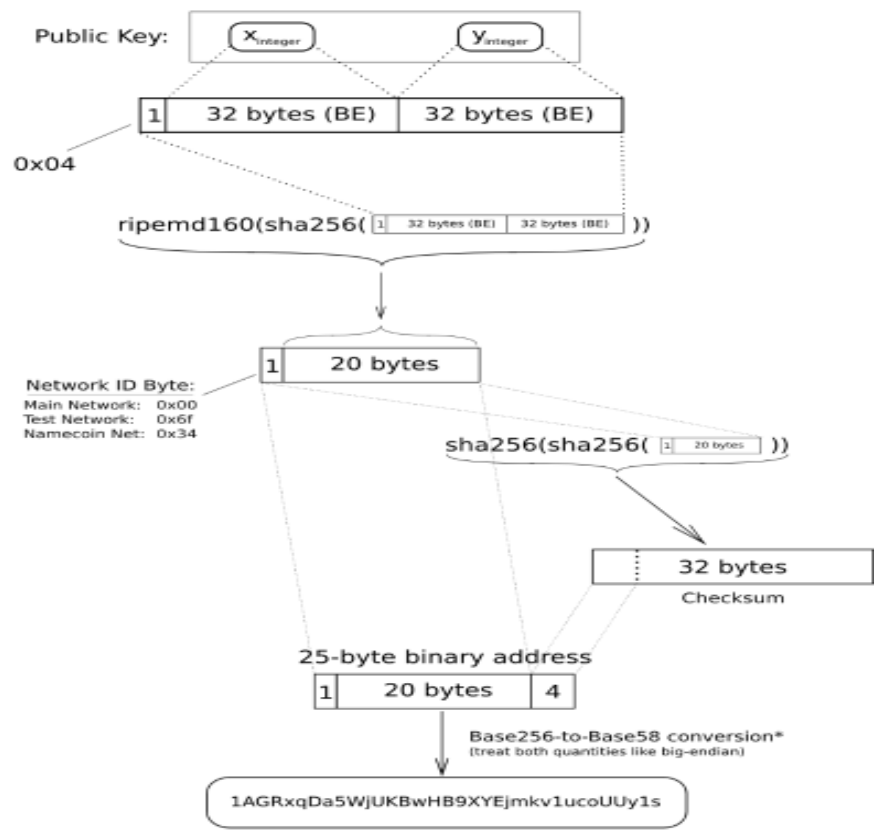

Figure 1. Public Key to Bitcoin Address

The proof-of-work system in the Bitcoin scheme is based on cryptographic hashing. The fixed size block header is first hashed using SHA-256 algorithm and the output is then hashed again using

SHA-256. The output number which is 256 bits long needs to be smaller than a specific target value which determines the difficulty of finding such an output value. The miner has to append a random number to the new block and make sure the hash value of new block begins with a series of zeros which can be done by brute-force. In the process of block chaining, the verification is done using previous user's public key and the miners use their private key for signing the block which they create or mine.

## 2. Background

### 2.1 Making of Bitcoins

The idea of digital or electronic currency is not new to the security and cryptography research communities. Over three decades before (in the year 1982), David Chum introduced the concept of digital currencies [5] . Similar to the use of paper cash, the digital currency described in this paper has three entities – Seller, Buyer and the bank (Refer Figure 2). As per the research study done in [6], [7] , any digital currency system should ideally support the following security measures:

- *Security* – The digital currency should not be forged, reused illegally.
- *Privacy* - Nobody including the bank should be able to reveal the end entities (Buyers/Sellers) and the products that they opt to choose. *Untraceablity*, *anonymity* and *unlinkability* are the sub-properties associated with privacy.
- *Transferability* - Without the involvement of the trusted third party (bank) the digital currency should be transferred between different customers, before it is stored in any of the customer's account.
- *Off-line Payment*- When the buyer buys some product or service from the seller, the seller should be able to verify the validity of the digital currency without an on-line inquiry with the bank.
- *Divisibility* - The end entities (Seller and Buyer) should be capable of subdivide the whole-sum digital cash into chunks of minute amount, without the aid of the bank.

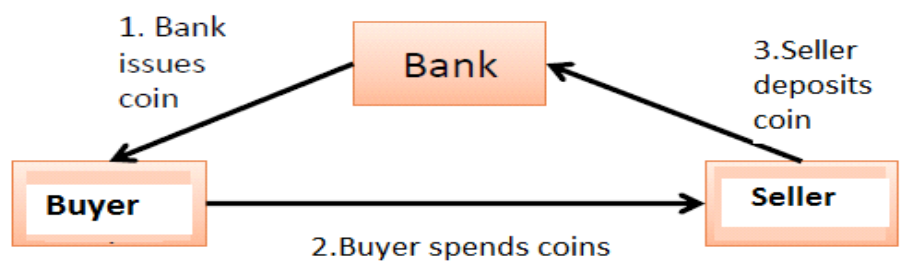

Figure 2 : The Framework of Digital Currency

Keeping Figure 2 in mind, we will now discuss the ideas that evolved in the process of making a digital currency [8].

**Idea 1: Blind Signature**
→ Transactions between buyer and the bank:
- Bank has an RSA signature key pair key (*e,d,n*) for signing 1€ coins (and different keys for 10€, 100€,...).
- The buyer creates a coin from random "serial number" *SN* and redundant padding required for RSA signature.
- Buyer generates a random number *R*, computes *coin · Re mod n*, and sends this to the bank.

- Bank computes *(coin · Re)d mod n = coin ·d · R mod n* and sends this to Buyer.
- Buyer divides with *R* to get the signed coin : *coin ·d mod n*.

→ As we can see the bank has signed the coin without seeing it and it cannot link the coin to buyer.
→ Now the buyer can pay 1€ to the seller by giving him the coin.
→ The seller deposits coin to bank; bank checks signature and accepts the same coin only once.

Problem at this point is *double spending*. Here the customers are anonymous; the cash is not physical, the bank cannot trace the transactions between the buyers and seller. If someone pays the same coin to two merchants, the described model does not provide an efficient way to avoid double payment.

**Idea 2: Double-spending detection**
→ Buyer must set "serial number" *SN = h( h(N) | h(N xor "Buyer") )* where *N* is a random nonce, *h* is a hash function, *Buyer* is the ID of the buyer.
→ After the buyer has given the coin to the seller, the seller asks the buyer to reveal one of *h(N), N xor "Buyer"* or *N, h(N xor "Buyer")*.
→ If the buyer spends the coin twice, he reveals his name with 50% probability.
→ Make each 1€ coin of k separately signed sub-coins, where detection probability $p = 1-2^{-k}$.
(Note : Coins will be quite large: k=128 with 2048-bit RSA signatures makes 32kB/coin)

Problem at this point is that how to enforce the buyer to create the "serial number" *SN* in the way described above. This approach fails to get the banks verification of the content of the messages signed blindly.

**Idea 3: Cut and Choose**
→ The buyer creates k pairs of sub-coins for signing.
→ Then the bank asks the buyer to reveal *N* ( a random nonce) for one sub-coin in each pair and signs the other one. (Note : cheating detection probability $p = 1-2^{-k}$)
→ Now the buyer can make anonymous payments but will be caught with probability $p = 1-2^{-k}$ if he tries to create an invalid coin or spend the same coin twice.

The Bitcoin, which evolved over time, to give a solution to problem with the above mentioned ideas strongly developed a concept of digital currency. As described earlier, Bitcoin is a transferable digital currency based on hash functions and cryptographic signatures. It is based on the peer-to-peer network system removing the need of the "bank" mentioned in the above ideas. Bitcoin transactions (Figure 3) are the direct transactions between public key pairs. Each transaction record includes the following components:
- Input Information – Previous transaction details ; Payer's signature.
- Output Information - Payee's public key hashes ; Payment amount.

Note that the *previous transaction amount* must be greater than or equal to *Output*.

History of the transactions proves the identity of the owner of the money at the time of each transaction. A public transaction log preferably noted as "Transaction ledger" contains all previously made transactions including the signatures as a sign of verification. It is updated every 10 minutes on average and it is used to check against the loophole of double-spending. In more technical terms it is nothing but the block chain containing hash of the previous block and Merkle hash of new transactions. Thus the latest block in effect contains the hash of all transactions ever. Double spending detection depends on the transaction block chain mechanism as per the Figure 4. The client software always chooses the longest branch.

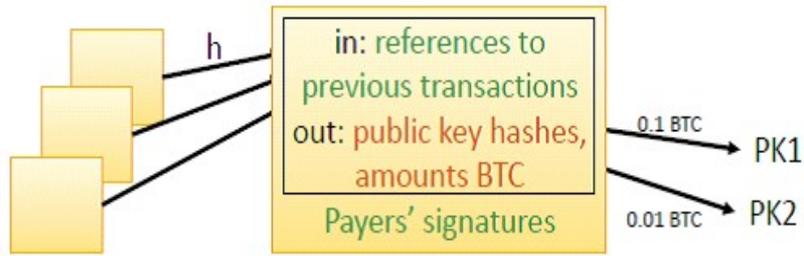

Figure 3 : Bitcoin Transaction.

After receiving the payment, the sellers publish the transaction to the P2P network and wait for 6 new blocks to include it (This is how a block gets officially recognized in the Bitcoin system). If someone controls more than 50% of the global hash rate, they can double spend but it is practically not feasible in the real time systems.

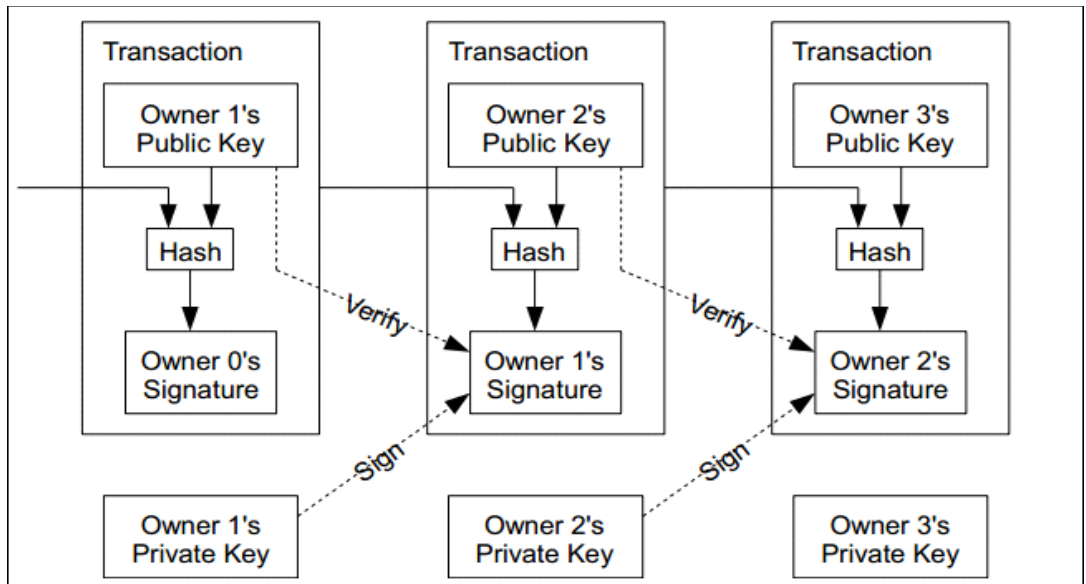

Figure 3 : Block Chaining in Bitcoins

## 2.2 Comparison with other digital currencies (A graphical representation and review)

In this section we review the present day statistics of Bitcoin and other digital currencies through graphical representation. Bitcoins has the clear advantage of having the longest coin span (Figure 5) as most of the other Bitcoin-like currencies are built based on it and are developed after the creation of Bitcoins with slight modification or as a variant.

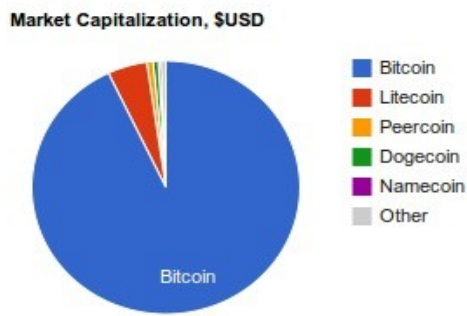

When we see the market capitalization of major digital currencies, we see that Bitcoin is a clear winner occupying almost 90% of the overall digital currency market with the current market capitalization (As of 30th April, 2014) of approximately 5.6 Billion US Dollars. Where as the next competitor in the list - Litecoin with 290 Million US Dollars clearly falling back the race.

Figure 4 : Market Cap of major digital Currencies [9]

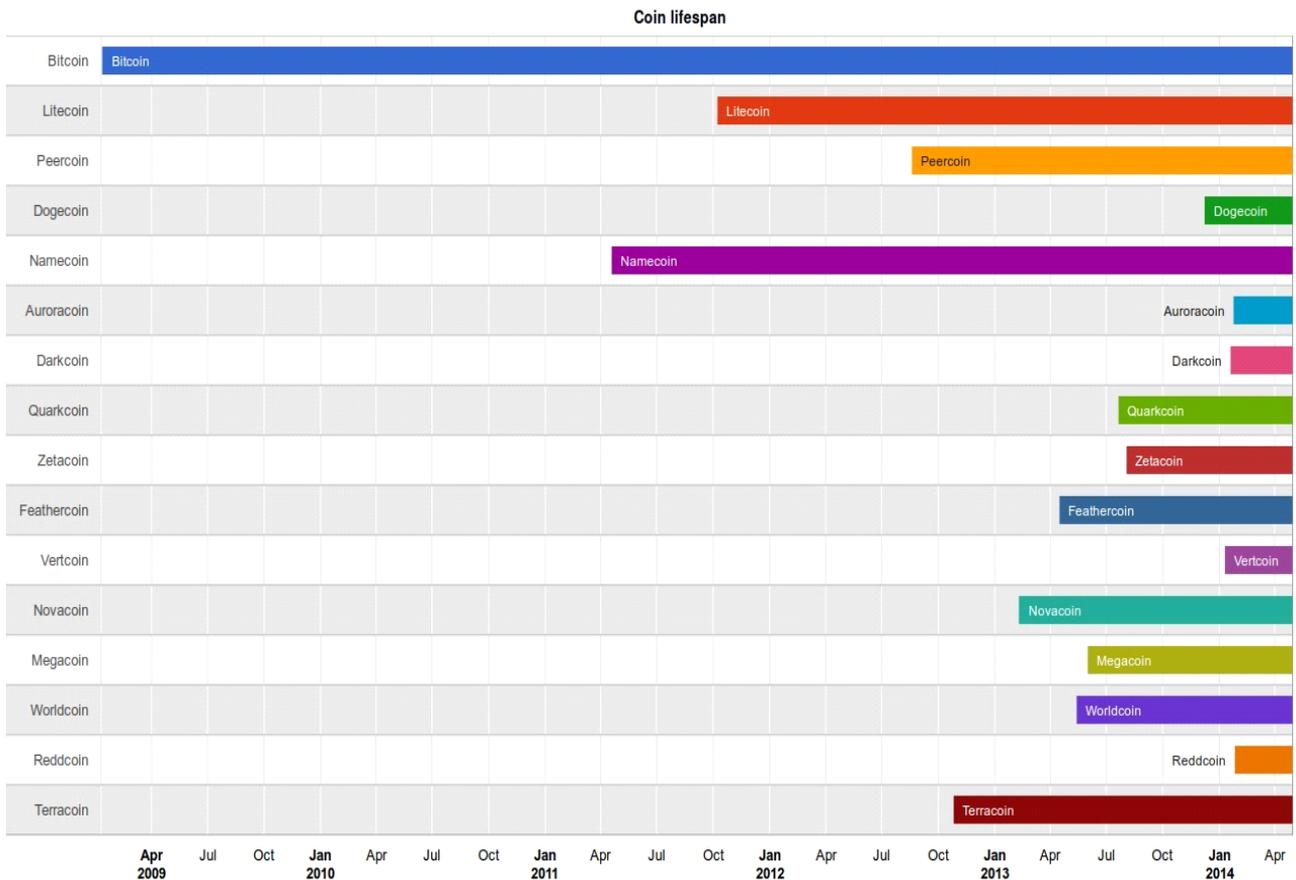

Figure 5 : Coin lifespan of major digital currencies.[9]

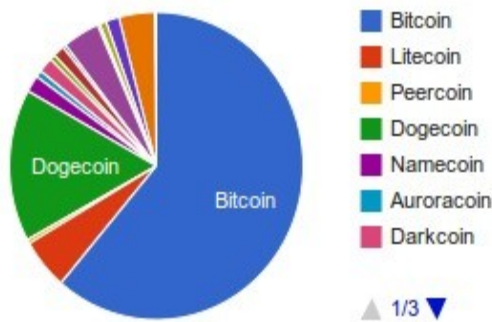

Figure 6 : No. of transactions in last 24 hours (30-05-14) [9]

When we have a look at the number of transactions happening on an average per day as per Figure 6, it is quite obvious that Bitcoin with largest user base is the being more active and hence making more than 50% of over all digital currency transactions in the world. As of 30th April, 2014, number of transactions of Bitcoin is approximately 65 thousand. The highest number of transactions happened in the history of Bitcoin so far per day is 102,010 during December, 2013.

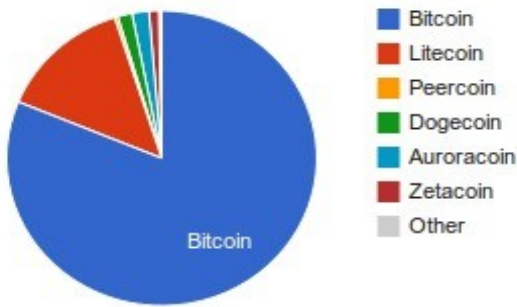

Figure 7: Digital coins sent in last 24 Hours ( 30th April, 2014) [9]

Figure 7 shows the statistics about the digital currencies sent in last 24 hours ( On 30th April, 2014). Again we have Bitcoins as winner with around 559,106 BTC ($248,674,391 USD) or 80% of overall digital currency being sent. Where as the second competitor Litecoin has just sent 4,119,512 LTC ($42,675,792 USD)

From the global digital currency market, we see from the above figures that following are the competitors [11, 12] of Bitcoins though they are lagging behind in the race by a predominantly large amount.

1. *Litecoin* -Litecoin is the favorite among "late comers", the investors who could not invest in the digital currency market during the initial stages of Bitcoin. It is forked from the Bitcoin system but the difference in Litecoin system is that it is crafted to be mined efficiently, cheaply, easily and faster than Bitcoin. It was first mined on First Mined Dec. 7, 2011.

2. *Peercoin* - Peercoin being another fork of Bitcoin, stands different from their parent by not having any limit on the amount of coins that can be created. Peercoin maintains a yearly inflation rate of 1% to increase the energy efficiency and to adjust the scalability from a long time perspective. Mining of Peercoin is more efficient than Bitcoin but is difficult as it has added security features. It was first mined on First Mined Aug. 16, 2012.

3. *Namecoin* - Namecoin is built on the same grounds of Bitcoin but with in the traditional sense it is not a currency at all. It serves as an alternative Domain Name System (DNS), controlling the *.bit* domain outside the realm of ICANN (Internet Corporation for Assigned Names and Numbers) registry. Here the customers buy the *.bit* domains and the domain itself is added to the block chains(i.e. to the public ledger keeping track of who owns the Namecoins-domains). It was first mined on April 17, 2011.

4. *Primecoin -* As the name suggests, Primecoin based on prime numbers for their encryption. In an attempt to keep the Primecoin system alive, the network discovers new prime numbers. Since prime numbers are favorites among mathematical and scientific communities, it is favorite among the researchers.

5. *Quarkcoin -* Quarkcoin is the fastest to mine digital currency with 0.5574 minutes for a transaction to be confirmed in the global peer-to-peer network. Though it is one of the youngest among the digital currencies, it is the most fast growing and most secure one. It has 9 rounds of encryptions using 6 different encryption algorithms which undoubtedly makes it most secure. It was first mined on July 21, 2013.

Figure 8 tabulates the top ten digital currencies(based on available market supply) among the existing 250 globally recognized digital currencies.

| # | Name | Market Cap | Price | Available Supply | Volume (24h) | % Change (24h) | Market Cap Graph (7d) |
|---|------|-----------|-------|------------------|--------------|----------------|----------------------|
| 1 | Bitcoin | $ 5,664,712,358 | $ 445.72 | 12,709,125 BTC | $ 16,769,976 | +0.66 % |  |
| 2 | Litecoin | $ 289,788,042 | $ 10.38 | 27,922,354 LTC | $ 3,380,418 | +2.80 % | 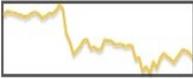 |
| 3 | Peercoin | $ 45,857,724 | $ 2.15 | 21,359,057 PPC | $ 132,925 | +3.47 % |  |
| 4 | Ripple | $ 40,689,033 | $ 0.005368 | 7,579,478,083 XRP* | $ 175,788 | +4.08 % | 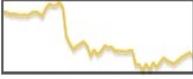 |
| 5 | Dogecoin | $ 37,359,054 | $ 0.000497 | 75,149,421,439 DOGE | $ 1,162,396 | +4.40 % |  |
| 6 | Mastercoin | $ 21,274,297 | $ 37.78 | 563,162 MSC* | $ 12,252 | +2.05 % | 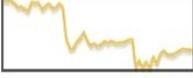 |
| 7 | Nxt | $ 21,111,694 | $ 0.021112 | 999,997,096 NXT* | $ 46,554 | -1.17 % |  |
| 8 | Namecoin | $ 18,188,809 | $ 2.11 | 8,632,332 NMC | $ 215,301 | +4.63 % | 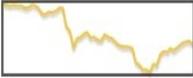 |
| 9 | BlackCoin | $ 9,408,134 | $ 0.13 | 74,518,504 BC* | $ 248,748 | -2.49 % |  |
| 10 | MaidSafeCoin | $ 8,023,016 | $ 0.017728 | 452,552,412 MAID* | $ 16,887 | -5.28 % | 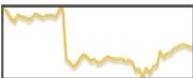 |

Fig 8: Cryptocurrency Market Capitalizations [ 10]

## 3. Success factors of Bitcoin

From the previous section we clearly see that Bitcoin with the unsaid advantage of being the oldest(early mined) digital currency deployed, being the global market winner. In this section we analyze the factors that influenced the success of the Bitcoin as per the paper [13].

- *No need for trusted third party(Bank) or central point :*
  As discussed earlier Bitcoin system does not need a bank or central authority to keep track of the transaction ledger, instead it relies on the trusted peer-to-peer network to do the same. Everyone in the network keeps a copy of the public ledger and this ledger is updated every 10 minutes on an average. Bitcoins system depends on a voting mechanism (for the next block to be on the top of the stack) from the peer-to-peer network users to avoid double-spending and resolve disputes.

- *Incentives for participation:*
  The Bitcoins economy systems is crafted in such a way that every fair user participation ensures monetary benefits (in the form of Bitcoins). So every Bitcoin user gets the Bitcoin which he successfully mines and adds to the public block-chain by exploiting his computer resource to solve the cryptographic puzzle. Miners receive the optional transaction fee which is very low at the moment. Unlike any other hard currency resource such as gold, Bitcoin mining cannot be done once there are 21 Million Bitcoins in the global network. It is estimated that around the year 2140, the limit of 20 Million Bitcoins will be reached and after that the participants will only be paid through transaction fee and not by mining.

- ***Bitcoin money supply which is predictable:***
  As the number of Bitcoin users increases, the difficulty of computational puzzle that they have to solve for the mining of Bitcoin increases. This ensures that the new coins are mined at a fixed rate with the growth of Bitcoin users. Hence early miners have an unstated advantage of mining the coins easily.

- ***Open Source code and Easily implementable modules:***
  Bitcoin is launched as an open source project which increased its flexibility with the users from various backgrounds. Because of its open source nature, more people indulged themselves to test, attest and participate in the Bitcoin community by creating various readily available implementation modules or applications for desktop and mobile computers.

- ***Support for scripting:***
  Though the feature of including the scripts with Bitcoin transactions is not utilized to its full extent, scripts for the following use cases [ 14,16 ] are doable at the time of writing this paper.
  → Standard generation/transaction to Bitcoin address .
  → Standard generation/transaction to IP address ( automatically getting the recipient to generate a Bitcoin address as long as their IP is known) .
  → Transaction with a message .
  → Hidden recipient address.

  Also in near future, we probably might see scripting for the following tasks[14,16] :
  → Providing a refundable deposit .
  → Escrow and dispute mediation .
  → Assurance contracts .
  → Using external state
  → Trading across chains

- ***Irreversibility of transactions:***
  Once the Bitcoin block is added to the block-chain, the transaction becomes irreversible. This is preferable for vendors who are hesitant to make business because of credit card fraud scams and charge backs. This feature helps many vendors to extend their business without the hesitation of usual monitory scams in the digital world. Linking the previous transactions in the Bitcoin system is inevitable which makes it resistant to double-spending without harming the anonymity of the end customer.

- ***Low fees of transaction:***
  At the moment there is an optional transaction fee for verifying the Bitcoin blocks through voting. But its optional and chosen by the payer. But after certain pint the transaction fees will be as profitable as the mining once the number of Bitcoins in circulation is reaching its upper limit.

All these factors have made Bitcoin to lead the digital currency market. In the long run, with the scope for future growth and promising tomorrow, Bitcoin is expected to serve as a stable digital currency.
# 4. Current challenges with Bitcoins

In this section we discuss the obstacles that can make a hindrance to Bitcoin in becoming the most popular currency in the long run. We first analyze the market trends from past 1 year, which will help to analyze the real world problems, then we will discuss the problems or challenges that have made the Bitcoin a subject to strong fluctuations.

**4.1 Various activities that affected the rise and fall of Bitcoin rate in the global market**

Figure 9 and 10 displays the fluctuation of Bitcoin rates in the year 2013 for two phases : January 2013 – May 2013 and October 2013 – December 2013 ( Since there was no much fluctuations due to public activities in the period July -September, that part has been omitted).

The factors that affected the face value of Bitcoins can be summarized as follows :
- Black hat communities hacking the Bitcoin systems
- Government restrictions on the digital currency transactions.
- A specific government issues new laws about the Bitcoin transactions
- Major digital market ( such as Silkroute) shutting down by the government.
- Malware attacks on the Bitcoin systems.
- Accidental loss of Bitcoins (Lack of measures to back up)

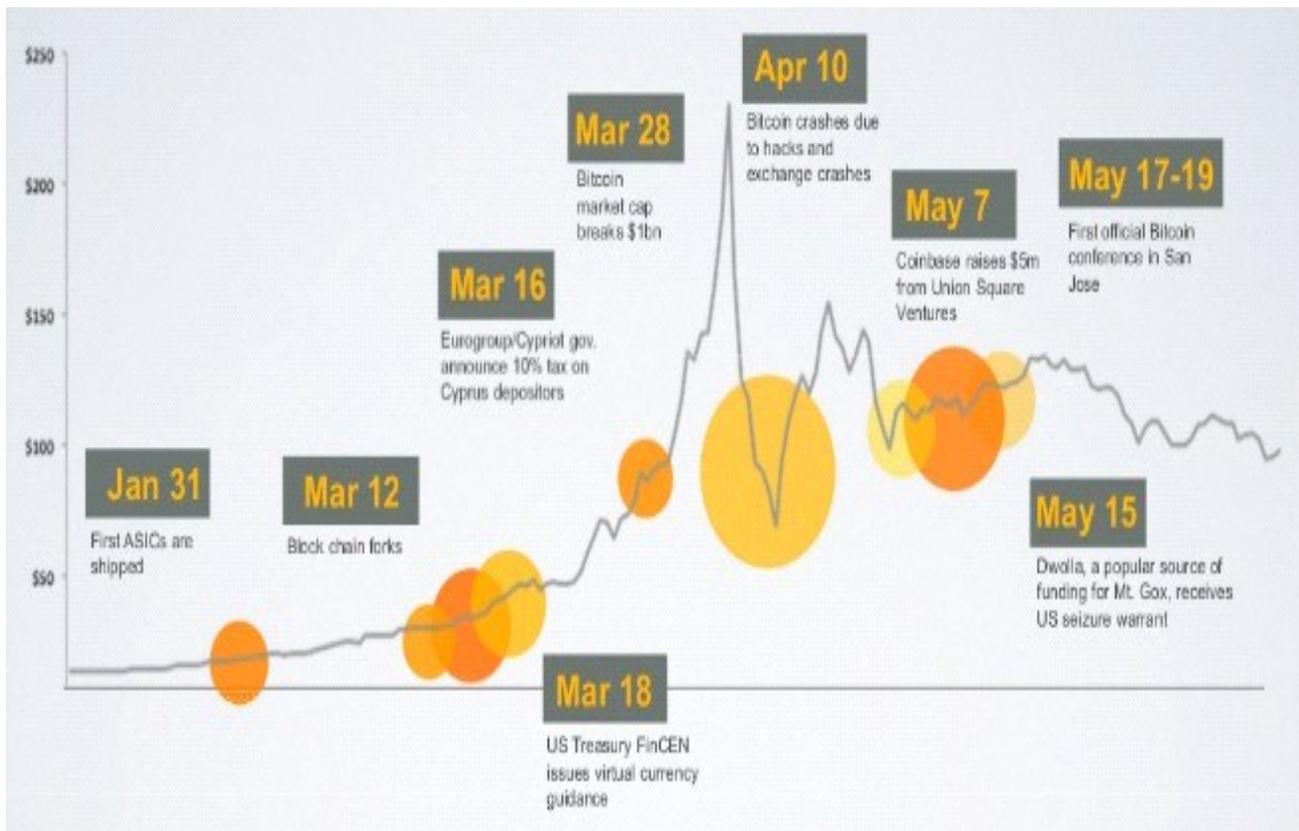

Figure 9 :  Bitcoin rates in the year 2014 (from January 2013 – May 2013) [17]

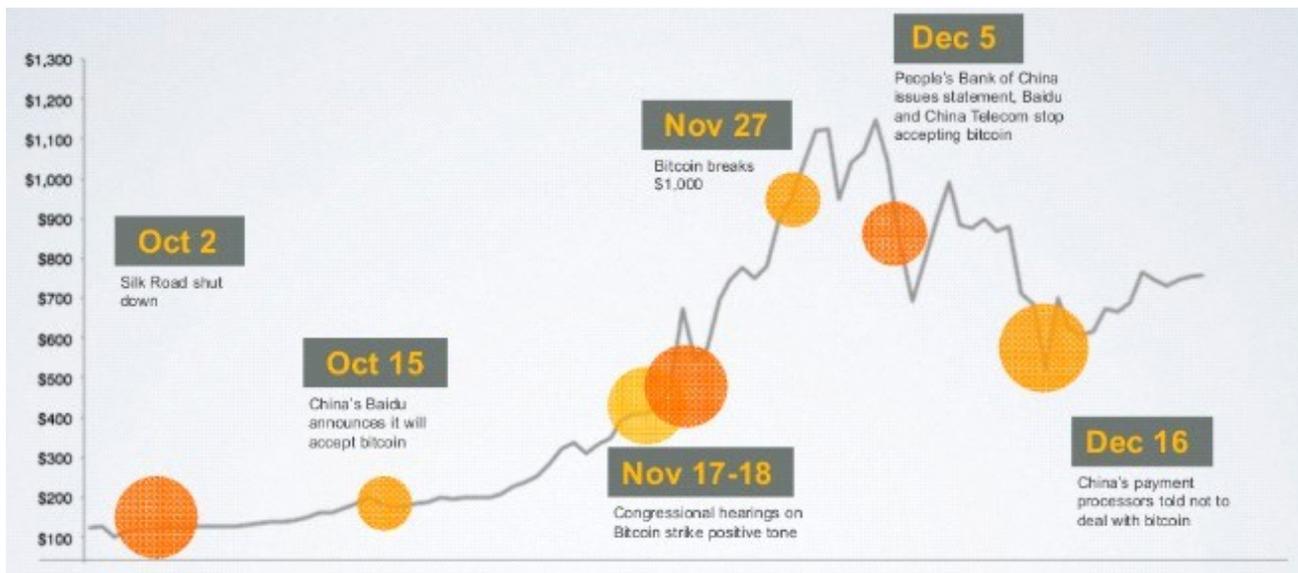
Figure 10 :  Bitcoin rates in the year 2014 (from October 2013 – December 2013) [17]

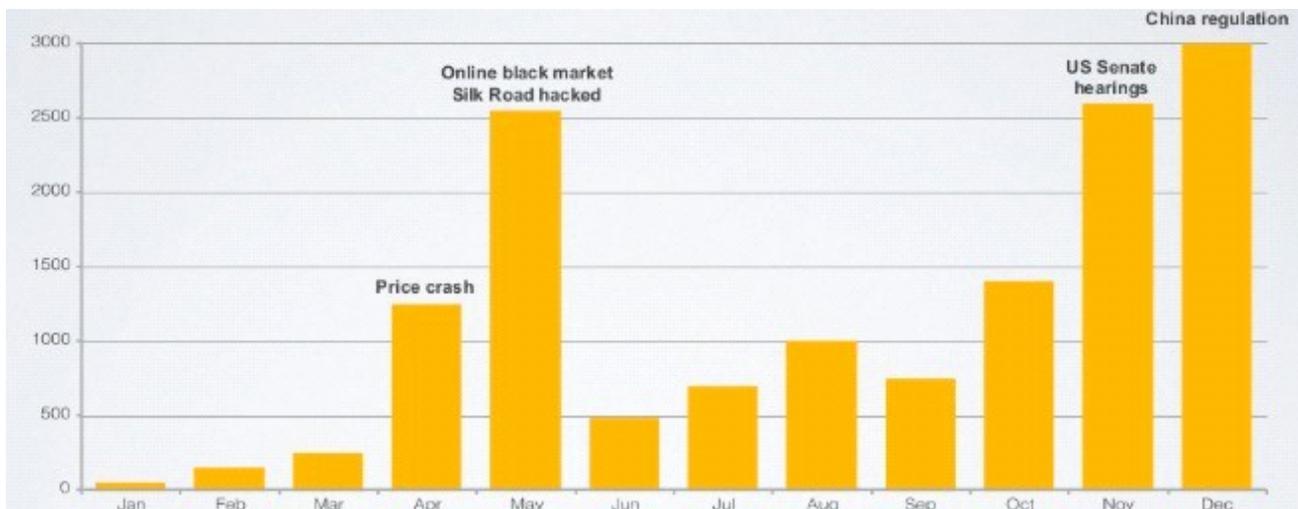
Figure 11 : Bitcoin media mentions in the year 2013 [18]

Another factor which has increased the number of transactions and face value of Bitcoins is media coverage. Because of media coverage, there has been clear unusual fluctuations in the Bitcoin transactions. Figure 11 shows the number of mentions in media about Bitcoins in the year 2013.

**4.2. Malware attacks and accidental loss of Bitcoins**

Although the Bitcoins system is protected by strong cryptographic mechanisms, there have been a lot of attacks on the system stealing millions of dollars by exploiting the loopholes in the system especially the private key storage mechanisms in Bitcoins. Thefts have been carried out by hacking the marketplaces or Bitcoin exchanges which can be quoted as *mass thefts*. Individual user's wallets have been stolen via *cryptocurrency Stealing Malware (CCSM)* and *Remote Access Trojans (RAT)* [19]. According to [19] we can classify the cryptocurrency Stealing Malware into categories namely: *Wallet Stealer*, *Credential stealer, Man-in-the-middle malware*, *RPC Automation malware.*

In the case of *wallet stealer*, the malware searches for the *wallet.dat* file or other known locations where the  wallet software keys are stored; with the help of key-logger or clipboard monitor, the malware fetches the passphrase for such protected file and send it to the attacker. The category of *Credential stealer,*  is a variant of *wallet stealer,* where it steals the credentials from a web based wallets ( such as Bitcoin exchanges). These kind of malware usually bypasses the security measures provided by the Bitcoin exchanges for authorized access such as two factor authorization with one-time passwords. In the *Man-in-the-middle* attacks, the malware alters the address of the recipient of a transaction before it gets signed. It then runs an observed sample in the background and monitors the contents of the clipboard. If such data is valid for a Bitcoin address, the malware replaces it with its own Bitcoin address to which victim will send Bitcoins without his knowledge. Usually the Bitcoin softwares include functionality called *Remote Procedure Call (RPC)* which establishes an interaction with other programs. In most of the cases the *RPC automation attacks* happens such that the attacker connects to the client running on a local TCP port and can steals the balance from the Bitcoin wallet with two to three commands which are automated. Though it is a possible attacking strategy, so far this kind of malware activities in a big scale have not been reported.

The best possible solution to malware based attacks on Bitcoins system can be achieved using *Threshold Cryptography* techniques [ 20]. Here the private keys are split into random shares and each share is kept in multiple locations such as user's desktop, smart phones, on the cloud , etc. The attacker can gain the private key only when he has access to threshold number of shares (certain fixed number) which is practically infeasible. Another variant of threshold cryptography technique is the concept of *Super Wallets* [13] . The idea in super wallets is that, there exits a personal bank for every user called "super wallet" where most of the Bitcoins can be kept. This super wallet is spread across multiple locations using the techniques of threshold cryptography. Alongside the user carries a *Sub Wallet* on the smart-phone. Small amount of money can be spend using Sub wallets and it can be refilled where and when necessary. When a big amount has to be payed, the user uses super wallet using the threshold devices. This adds another layer of security to the threshold cryptography imposed Bitcoin systems.

Another major challenge is that the Bitcoins environment is facing is accidental loss of Bitcoins either because of system failures or human errors which turns the Bitcoins into *Zombies.* One of the possible solutions to this is taking backups of the wallets and keeping them safe like any other cryptographic assets. But the problem with this approach is that the wallet file keeps growing. To avoid that one can generate private key not randomly but using a Pseudo Random Generator from a master secrete which would not change. Other solutions could be using password based encryption of the wallet files. Users can opt to choose multiple passwords for various tasks (e.g. everyday password, big transaction password, etc.). These approaches can be collaborated with dedicated hardware devices called Trusted-path Devices which let the human inputs and cryptographic data output safe from any malware.

### 4.3. Privacy issues

Due to the decentralized and distributed nature of Bitcoins, everyone can obtain the history of transactions without much effort. Transactions in the Bitcoin like systems can be visualized as Directed Acyclic Graphs (DAG) with each vertex in graph representing a single transaction. Study of such Directed Acyclic Graphs [21] resulted in finding that the rich users (all large transactions are related to a particular Bitcoin address and to a single transaction done in November 2010) along with the clear result about the typical user behavior about spending and acquisition of Bitcoins in the network. Few such results deduced are as follows :  There are 98% of the addresses which have fewer than 10 Bitcoins;  47% of the general transactions make less than 0.01 Bitcoins; 84% of the transactions are less than 10 Bitcoins.

On the other hand, analysis of a user network would also lead to many results which contradicts the privacy of the Bitcoin system. In the present day Bitcoin system, there is no possibility that several users together can pay for one payment. So whenever a *multi-input payment* happens(that happen when value owned by a particular address is less than the amount required for payment), it means all the addresses belong to a same user. Similarly whenever the transaction has two output addresses with one being the old address (the address that has appeared in the previous transaction log) and one new address, then it is quite obvious that the new address is the *shadow address* of the user who is paying. Privacy can also be leaked by the TCP/IP layers. Using the mapping features in the TCP/IP protocol, one can easily map out and check the geographical locations of the Bitcoin transactions being made. If the IP addresses belong to far away regions it could reveal that the addresses could belong to two different users. Most of the times the companies which accepts the Bitcoin might require the email address and some sort of personal Identification. If that is revealed, it is a complete breach of privacy in the Bitcoin system. Probable solution to the privacy related challenges can be usage of *Mixtures* and *Fair Exchange Protocol*, which will be discussed in Section 5.

**4.4. Scalability Issues**

Data retentions and communication failures are one of the major challenge in the Bitcoin networks where broadcasts of transactions and blocks in a timely manner matters the most. Filtering of Bitcoin nodes as *Clients* and *Verifiers* based on the bandwidth, computational power and battery supply could resolve the scalability issues. Verifiers mine new Bitcoins, and the desktop computers of the users with more computational resources. Clients on the other hand are mainly interested in spending the coins which a particular user has and hence it can be his smart-phones. This type of filtering helps to improve the scalability issues in Bitcoins such as, authenticity of all blocks and transactions will be verified cryptographically by all the Bitcoins nodes as soon as it relieve them. One such filtering service can be provided using a trusted third party cloud service which filters all Bitcoins transactions and send it to registered client or verifier by determining the transaction is payable to one or more of its public keys [13]. It is important that Bitcoin filtering services to support features like *Unlinkability without the capability*, *Forward Security* and *reasonable false positive – low false negative* strategy.

One of the other major problems with Bitcoins ecosystem is *delay associated with transaction confirmation* which is usually 10 minutes for the newly generated block to get confirmed in the global transaction block chain. This is a major drawback when the user has to pay Bitcoins to a system where timing is critical(Such as Stock Exchanges or On-Demand Playback), where the chances of double spending could be high. One probable solution is introducing a *semi-trusted bank* as in intermediate which is capable of issuing Bitcoin equivalent of checks of the cashier. Another feasible approach is to reduce the block confirmation interval to 10 seconds from 10 minutes by adjusting the computational puzzles which are used during mining.

Growing size of private key storage is another issue associated with the Bitcoin ecosystem though it helps to achieve better anonymity. The solution for this problem is using *Pseudo Random Generator* for generating private keys(as discussed in section 4.2) and associating *expiration dates* for public-keys.

**5. Achieving balance between anonymity and trust**

Anonymity of the users in Bitcoin ecosystems is based upon pseudonyms which are nothing but the Bitcoin addresses. It is possible that any user may own more than one Bitcoin address. The Bitcoin enforces anonymity and unlinkability by allowing users to have different addresses and public keys in every transaction. However as discussed in section 4.3 and in [22] it is possible to link the users and hence privacy of the concept will be breached. Whenever there is more anonymity to be achieved, it reduces the trust in the network as it they are indirectly proportional. Having a right balance between anonymity and trust is very much important in such digital currencies is very much essential. This balance can be achieved with the help of *Mixers* and *Fair Exchange Protocols*.

*Mixtures* are the third party trusted services which was improvised to improve the anonymity of the system. It collects Bitcoins from the users, randomly mixes them and after which Bitcoin with same denomination will be returned to the users[13]. But it does not go hand in hand with the Bitcoin's policy of *Trust No One.* Because if there exists a malicious mixer, it might refuse to pay back the Bitcoins and it might collects all the private information of the user. So all the advantage brought by this feature can be lost.

To avoid the compromising nature of mixers, we can foresee *Fair Exchange Protocol* as a better alternative. This approach is works on a backward compatible manner where two users can exchange Bitcoins without third party intervention. Also the probability that a user will cheat is less. The fair exchange protocol consists of three types of transactions namely *Commitment Transaction*, *Refund Transaction* and *Claim Transaction*. It also goes through three different phases namely *Secret Setup Phase*, *Transaction Setup phase* (Separately for two end users) and *Money Claim Phase*. In the secret set up phase, two end users using this protocol performs key generation and exchange public keys and later they use these keys for different transactions which ensures unlinkability. In the transaction set up phase, both users take turns to establish a minimal trust relationship by verifying each other's signatures and at the end committing to exchange the Bitcoins as it happens in mixers. Along with this each of them will set a *lock time* to make ensure the timely exchange of Bitcoin denominations. In the money claim chase, the end users can claim only those which is affiliated to them by changing the lock time to current time. The fair exchange protocol can be amalgamated with untrusted mixers without leaking the privacy (as depicted in figure 12).

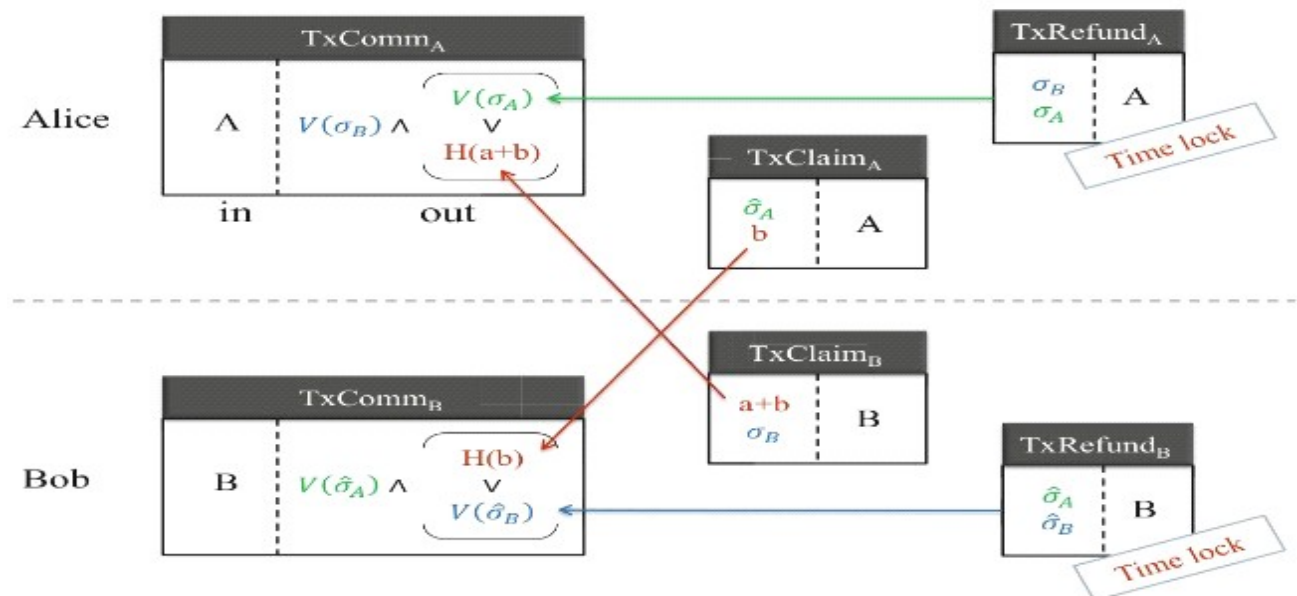

Figure 12: Fair Exchange Protocol : Mixing Bitcoins with an untrusted mixer [13]

## Conclusion

The concept of Bitcoins has revolutionized the whole technology and financial world. Analyzing the current market trends in the digital currency world revealed that Bitcoin is the winner among all existing and newly forking digital currencies. The success attained by Bitcoin on various factors makes it as the market leader among the digital currencies. Alongside, analyzing the ups and downs of the market value of Bitcoins in the past year implicates the probable flaws in the system and hence showcasing the solutions that could be implemented to overcome those flaws. Assuming that these solutions are implemented successfully, it gives a positive hope about Bitcoin revolution to be continuing for a long run, providing economical benefits preserving the privacy of people around the globe without depending on any government or third party authorities.